\documentstyle[preprint,aps,epsfig,12pt]{revtex}

\begin{document}

\begin{flushright}
  UWThPh-1998-10 \\
  HEPHY-PUB 689/98 \\
  hep-ph/9805248
\end{flushright}

\begin{center}
{\Large \bf Sfermion Pair Production at {\boldmath $\mu^+ \mu^-$} Colliders}
\end{center}

\begin{center}
\large{A. Bartl$^1$, H. Eberl$^2$, S. Kraml$^2$,
        W. Majerotto$^2$, W. Porod$^{1*}$}\\

\small{\it
$^1$Institut f\"ur Theoretische Physik, Universit\"at Wien,
    A-1090 Vienna, Austria \\
$^2$Institut f\"ur Hochenergiephysik der \"Osterreichischen
            Akademie der Wissenschaften, A-1050 Vienna, Austria}
\end{center}

\footnotetext[1]{Present address:
                 Instituto de F{\'\i}sica Corpuscular -- IFIC/CISC,
                 Departament de  F{\'\i}sica Te{\'o}rica,
                 E-46100 Burjassot, V{\'a}lencia, Spain}
                 
\begin{abstract}
We discuss pair production of stops, sbottoms, staus and tau--sneutrinos at a 
$\mu^+ \mu^-$ collider. We present the formulae for the production cross
sections and perform a detailed numerical analysis within the Minimal 
Supersymmetric Standard Model. In particular, we consider sfermion production 
near $\sqrt{s} = m_{H^0}$ and  $\sqrt{s} = m_{A^0}$.
\end{abstract}

\section{Introduction}

The search for Supersymmetry (SUSY) \cite{susy,kane} is one of the main 
issues in the
experimental programs at LEP2 and TEVATRON. It will play an even more important
r\^ole at the future colliders LHC, $e^+ e^-$ linear colliders with
an energy range up to 2~TeV, and $\mu^+ \mu^-$ colliders with an energy range
up to 4~TeV.  TEVATRON and LHC are well--designed to discover SUSY. At these 
machines even some precision measurements are possible \cite{paige97}. 
However, for a precise determination of the underlying SUSY parameters
lepton colliders will be necessary. Owing to their good energy resolution
$\mu^+ \mu^-$ colliders are well suited for this purpose 
\cite{gunion,Gunion96}. The most exciting feature is the possibility of
producing Higgs bosons in the s--channel \cite{gunion,Gunion96,haber}.

In this paper we study the production of third generation sfermions
in $\mu^+ \mu^-$ annihilation, paying particular attention to the energy 
range near the Higgs boson resonances. 
Our framework is the Minimal Supersymmetric Standard Model 
(MSSM) \cite{kane,Gunion86}.
The MSSM implies the existence of five physical
Higgs bosons: two scalars $h^0, \, H^0$, one pseudoscalar $A^0$, and two 
charged ones $H^{\pm}$ \cite{Gunion86,higgshunter}. Every Standard Model (SM) 
fermion
has two supersymmetric partners, one for each chirality state denoted by
$\tilde f_L$ and $\tilde f_R$.

The sfermions of the third generation are particularly interesting 
\cite{Ellis83,Bartl97,desy} because their phenomenology is different compared 
to that of 
the sfermions of the first and second generation. The reasons for this are the 
mixing between $\tilde f_L$ and $\tilde f_R$ and the large Yukawa couplings.

In particular, the top quark and the stops give
substantial contributions to Higgs boson masses due to radiative corrections
(see e.g. \cite{ellis,haber91}). 
Moreover, the top and stop contributions to the 
renormalization group equations play an essential r\^ole in inducing 
electroweak symmetry breaking, when the Higgs parameters evolve from the GUT 
scale to the electroweak scale \cite{ross}. Therefore, the couplings of the 
stops to the neutral Higgs bosons are of special interest. Also the tau and
bottom Yukawa couplings can be large if $\tan \beta$ is large, giving
rise to similar effects in the phenomenology of sbottoms and staus.

The paper is organized in the following way: In Sec.~II we present the
underlying parameters and give the formulae for the production cross sections.
In Sec.~III we discuss numerical results for sfermion pair
production. In Sec.~IV we summarize the main results. In the Appendix we
give the Higgs sfermion couplings.

\section{Sfermion Pair Production}

In this section we present the formulae for sfermion mixing and the
production cross sections of sfermions at a $\mu^+ \mu^-$ collider.
The main parameters for the following discussion are $m_{A^0}$,
$\mu$, $\tan \beta$, $M_{\tilde D}$, $M_{\tilde Q}$, $M_{\tilde U}$, 
$M_{\tilde E}$, $M_{\tilde L}$, $A_b$, $A_t$, and $A_\tau$
\cite{kane,Gunion86}. $m_{A^0}$ is the mass of the pseudoscalar Higgs boson,
$\mu$ is the Higgs mixing parameter in the superpotential, and 
$\tan \beta = v_2 / v_1$,
where $v_i$ denotes the vacuum expectation value of the Higgs doublet $H_i$.
$M_{\tilde D}$, $M_{\tilde Q}$, $M_{\tilde U}$, $M_{\tilde E}$ and 
$M_{\tilde L}$ are soft SUSY breaking masses
for the sfermions, $A_{b}$, $A_{t}$, and $A_\tau$ are trilinear 
Higgs--sfermion parameters.

The mass matrix for sfermions in the $(\tilde{f}_L,\:\tilde{f}_R)$ basis
has the following form:

\begin{eqnarray} \label{sqm}
  {\cal M}^2_{\tilde{f}} = \left(\begin{array}{ll}
                       m_{\tilde{f}_{L}}^{2} & a_{f} m_{f} \\
                       a_{f} m_{f}  & m_{\tilde{f}_{R}}^{2}
                   \end{array} \right)
\end{eqnarray}
with
\begin{eqnarray}
  m_{\tilde{t}_{L}}^{2}&=& M_{\tilde Q}^2 + m_t^2 + m_Z^2\cos 2\beta\,
  (\textstyle \frac{1}{2} - \frac{2}{3} \sin^2\theta_W ) , \nonumber \\
   m_{\tilde{t}_{R}}^{2}&=&M_{\tilde U}^2 + m_t^2 + \textstyle \frac{2}{3}
                        m_Z^2\cos 2\beta\,\sin^2\theta_W , \nonumber \\
  m_{\tilde{b}_{L}}^{2}&=&M_{\tilde Q}^2 + m_b^2 - m_Z^2\cos 2\beta\,
    (\textstyle \frac{1}{2} - \frac{1}{3} \sin^2\theta_W ) , \nonumber \\
  m_{\tilde{b}_{R}}^{2}&=&M_{\tilde D}^2 + m_b^2 -
      \textstyle \frac{1}{3} m_Z^2\cos 2\beta\,\sin^2\theta_W , \nonumber \\
  m_{\tilde{\tau}_{L}}^{2}&=&M_{\tilde L}^2 + m_\tau^2 - m_Z^2\cos 2\beta\,
    (\textstyle \frac{1}{2} - \sin^2\theta_W ) , \nonumber \\
  m_{\tilde{\tau}_{R}}^{2}&=&M_{\tilde E}^2 + m_\tau^2 -
      \textstyle m_Z^2\cos 2\beta\,\sin^2\theta_W ,
\label{smatrix}
\end{eqnarray}
and
\begin{eqnarray}
  a_t = A_t - \mu\, \cot\beta , \nonumber \\
  a_b = A_b - \mu\,\tan\beta , \nonumber \\
  a_\tau = A_\tau - \mu\,\tan\beta.
  \label{offdiag}
\end{eqnarray}

The mass eigenstates $\tilde{f}_1$ and $\tilde{f}_2$ are related to
$\tilde{f}_L$ and $\tilde{f}_R$ by:
\begin{eqnarray}
  {\tilde{f}_1 \choose \tilde{f}_2} =
    \left(\begin{array}{rr}
        \cos \theta_{\tilde f} & \sin \theta_{\tilde f} \\
       -\sin \theta_{\tilde f} & \cos \theta_{\tilde f} \end{array} \right)\:
    {\tilde{f}_L \choose \tilde{f}_R}
  \label{mixing}
\end{eqnarray}
with the eigenvalues
\begin{eqnarray}
  m_{\tilde{f}_{1,2}}^2 & = & \textstyle \frac{1}{2}\,
       (m_{\tilde{f}_L}^2 + m_{\tilde{f}_R}^2)
   \mp \frac{1}{2} \sqrt{(m_{\tilde{f}_L}^2 - m_{\tilde{f}_R}^2)^2
                           + 4\,a_f^2 m_f^2}. 
  \label{eq:masses}
\end{eqnarray}
The mixing angle $\theta_{\tilde f}$ is given by
\begin{eqnarray}
  \cos \theta_{\tilde f} = \frac{- a_f m_f}
           { \sqrt{(m_{\tilde{f}_L}^2-m_{\tilde{f}_1}^2)^2 + a_f^2 m_f^2}},
\nonumber \\ 
  \sin \theta_{\tilde f} = \sqrt{ \frac{(m_{\tilde{f}_L}^2
                                  -m_{\tilde{f}_1}^2)^2}
                   {(m_{\tilde{f}_L}^2-m_{\tilde{f}_1}^2)^2 + a_f^2 m_f^2}}.
 \label{eq:mixangl}
\end{eqnarray}
The mass of the tau--sneutrino is given by
\begin{equation} 
m^2_{\tilde \nu_\tau} = M_{\tilde L}^2 
                        - \textstyle \frac{1}{2} m_Z^2\cos 2\beta.
\end{equation}

\begin{figure}[t!] 
\begin{picture}(320,200)
\put(30,0){\mbox{\psfig{file=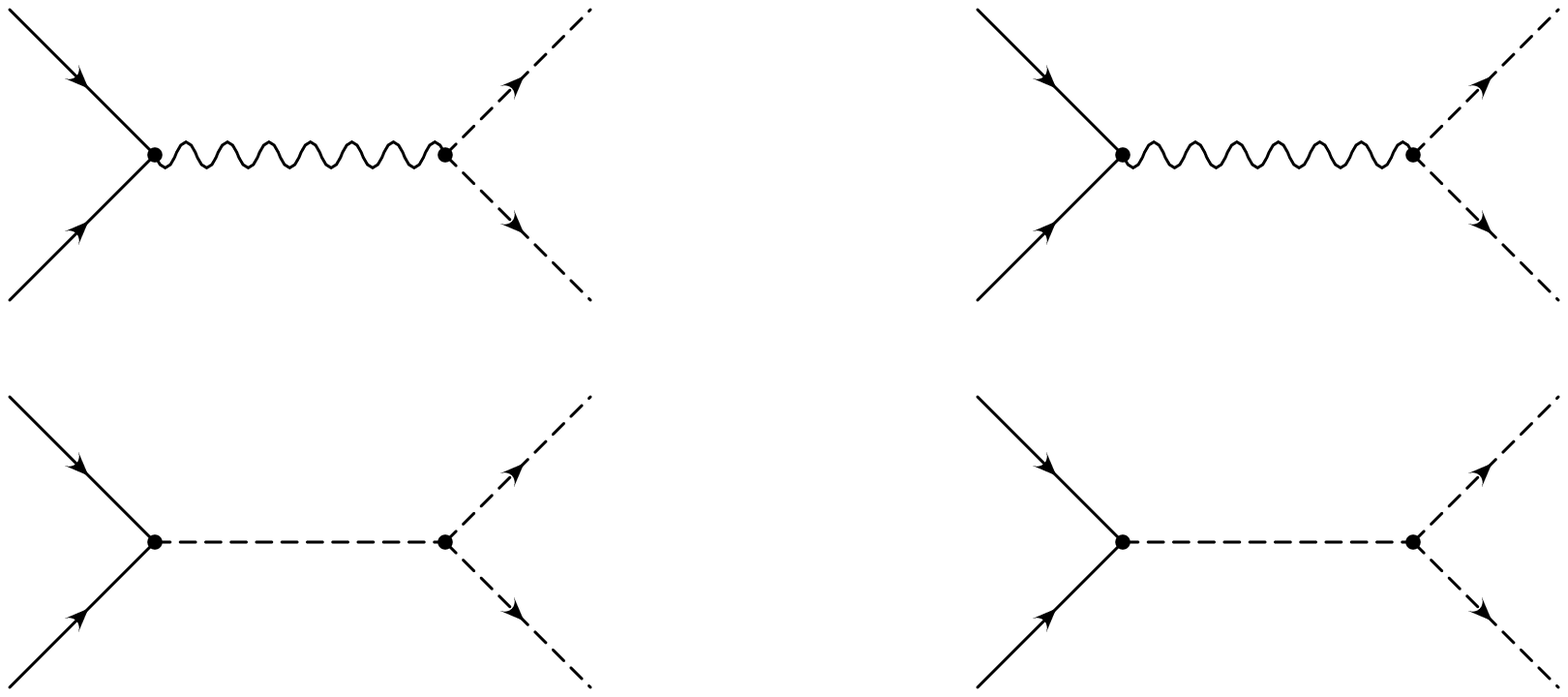,height=2.5in,width=4.7in}}}
\put(10,185){\mbox{a)} }
\put(45,172){\mbox{$\mu^-$} }
\put(45,112){\mbox{$\mu^+$} }
\put(80,150){\mbox{$\gamma, \, Z^0$} }
\put(135,172){\mbox{$\bar{\tilde f_i}$} }
\put(133,110){\mbox{$\tilde f_i$} }
\put(45,70){\mbox{$\mu^-$} }
\put(45,10){\mbox{$\mu^+$} }
\put(80,46){\mbox{$h^0, \, H^0$} }
\put(135,70){\mbox{$\bar{\tilde f_i}$} }
\put(133,8){\mbox{$\tilde f_i$} }
\put(222,185){\mbox{b)} }
\put(257,172){\mbox{$\mu^-$} }
\put(257,112){\mbox{$\mu^+$} }
\put(300,150){\mbox{$Z^0$} }
\put(347,172){\mbox{$\bar{\tilde f_2}$} }
\put(345,110){\mbox{$\tilde f_1$} }
\put(257,70){\mbox{$\mu^-$} }
\put(257,10){\mbox{$\mu^+$} }
\put(278,45){\mbox{$h^0, \, H^0, \, A^0$} }
\put(347,70){\mbox{$\bar{\tilde f_2}$} }
\put(345,8){\mbox{$\tilde f_1$} }
\end{picture}
\vspace{10pt}
\caption{Feynman graphs for sfermion production in $\mu^+ \mu^-$ 
         annihilation: a) for $\tilde f_i \bar{\tilde f_i}$ $(i = 1,2)$,
         b) for $\tilde f_1 \bar{\tilde f_2}$.}
\label{feynman}
\end{figure}

Figure \ref{feynman} shows the Feynman graphs for the processes 
$\mu^+ \mu^- \to \tilde f_i \bar{\tilde f_j}$ $(i,j = 1,2)$.  The total and 
the differential cross sections read up to ${\cal O}(m_\mu^2)$:
\begin{eqnarray}
\sigma(\mu^+ \mu^- \to \tilde f_i \bar{\tilde f_j})
& = & c_{ij}\left[\frac{4}{3}\frac{\lambda_{ij}}{s^2}\, T_{VV} +  
2 \, \frac{m_{\tilde f_i}^2-m_{\tilde f_j}^2}{s} T^a_{VH} +  2 \, T_{HH}
 \right] \, , \label{sigtot} \\
\frac{d \, \sigma(\mu^+ \mu^- \to \tilde f_i \bar{\tilde f_j})}
     {d \, \cos \vartheta}
q& = & c_{ij} \left[ \frac{\lambda_{ij}}{s^2}  T_{VV} \, \sin^2 \vartheta
           + \frac{m_{\tilde f_i}^2-m_{\tilde f_j}^2}{s} T^a_{VH}  +  
             \frac{\lambda^{1/2}_{ij}}{s} T^b_{VH} \cos \vartheta 
+ T_{HH} \right] 
\label{sigdiff}
\end{eqnarray}
with the kinematic function $\lambda_{ij} = 
(s - m_{\tilde f_i}^2 - m_{\tilde f_j}^2)^2
 - 4 m_{\tilde f_i}^2 m_{\tilde f_j}^2$,
$\vartheta$ the scattering angle of $\tilde f_i$, and 
\begin{eqnarray}
c_{ij} & = & \frac{\pi N_C \alpha^2}{4 \, s^2}\lambda^{1/2}_{ij} \, ,
\end{eqnarray}
where
$N_C$ is a colour factor which is 3 for squarks and 1 for sleptons.

$T_{V\!V}$ denotes the contribution from $\gamma$ and $Z^0$
exchange, $T^{a,b}_{V\!H}$ the interference terms between gauge
and Higgs bosons, and $T_{H\!H}$ the contribution stemming from the 
exchange of Higgs bosons. 
The pure gauge boson contribution, the first term of Eqs.~(\ref{sigtot}) and
(\ref{sigdiff}), is the same as for $e^+ e^- \to \tilde f_i \bar{\tilde f_j}$ 
\cite{eberl}. Notice that the gauge boson term shows a $\sin^2 \vartheta$ 
dependence whereas the terms proportional to $T_{H\!H}$ and $T^{a}_{V\!H}$ are
independent of $\vartheta$. The term proportional to $T^b_{V\!H}$ shows a 
$\cos \vartheta$ dependence giving rise to a forward--backward asymmetry. 
However, $T^b_{V\!H}$ is proportional to $m_\mu$ (see the following equations)
and, therefore, is rather small.

We obtain: \\
1) $i = j$
\begin{eqnarray}
T_{VV} & = & 
  e^2_f -  \frac{e_f a_{ii} v_\mu \, s}
                       { 8 \sin^2 \theta_W \cos^2 \theta_W} Re[ D(Z) ]
 +  \frac{a^2_{ii} (v_\mu^2+a_\mu^2) \, s^2}
         {256 \, \sin^4 \theta_W \cos^4 \theta_W }  | D ( Z ) |^2 \, ,
 \nonumber \\
T_{HH} & = &  \,  h_\mu^2\,   \frac{s}{2 e^2 \sin^2 \theta_W}
     \, |G^{h^0}_{ii} \sin \alpha D(h^0) - G^{H^0}_{ii} \cos \alpha D(H^0) |^2
 \, , \nonumber \\
T^a_{VH} & = &  0 \, ,\nonumber \\
T^b_{VH} & = &  h_\mu m_\mu \, \frac{2 \sqrt{2}}{e \sin \theta_W}
   \Big[ e_f Re[ G^{h^0}_{ii} \sin \alpha D(h^0)
                - G^{H^0}_{ii} \cos \alpha D(H^0) ]  \nonumber \\
& & - \frac{v_\mu a_{ii} \, s}{16 \sin^2 \theta_W \cos^2 \theta_W} 
  \, Re[D^*(Z) ( G^{h^0}_{ii} \sin \alpha D(h^0) 
               - G^{H^0}_{ii} \cos \alpha D(H^0) ) ] \Big] \, ,
\label{tii}
\end{eqnarray}
2) $i \neq j$
\begin{eqnarray}
T_{VV} & = & \frac{a^2_{ij} (v_\mu^2+a_\mu^2) \, s^2}
                  {256 \, \sin^4 \theta_W \cos^4 \theta_W } \, | D ( Z ) |^2
 \, , \nonumber \\
T_{HH} & = &  h_\mu^2 \, \frac{s}{2 e^2 \sin^2 \theta_W} \,
       \left[ | G^{h^0}_{ij} \sin \alpha D(h^0)
                - G^{H^0}_{ij} \cos \alpha D(H^0) |^2 
          + | \sin \beta \, G^{A^0}_{12}  D(A^0) |^2 \right] \, , \nonumber \\
T^a_{VH} & = &  - h_\mu m_\mu \,
      \frac{ \sqrt{2} \, a_{ij} \, a_\mu  \sin \beta \, \, \,s}
           {8 e \sin^3\theta_W \cos^2 \theta_W}
           G^{A^0}_{12} \, Re[ D^*(Z) D(A^0) ] \, , \nonumber \\
T^b_{VH} & = &  -h_\mu m_\mu\, 
     \frac{\sqrt{2} \, v_\mu a_{ij} \, s}{8 e \sin^4 \theta_W \cos^2 \theta_W} 
Re[D^*(Z) (G^{h^0}_{ij} \sin \alpha D(h^0) - G^{H^0}_{ij} \cos \alpha D(H^0) 
)],
\label{tij}
\end{eqnarray}
where $h_\mu$ is the Yukawa coupling of the muon, 
$h_\mu = g \, m_\mu /(\sqrt{2} \, m_W \cos\beta)$,
$D(i) = 1 / ( (s - m^2_i) + i \Gamma_i m_i)$;
$e_f$ is the charge of the sfermions ($e_t = 2/3, e_b = - 1/3, e_{\tau} = -1$) 
$v_\mu$ and
$a_\mu$ are the vector and axial vector couplings of the muon to the $Z$~boson,
$v_\mu = -1 + 4 \sin^2 \theta_W$, $a_\mu = -1$,
and $a_{ij}$ are the corresponding couplings $Z \tilde f_i \bar{\tilde f_j}$:
\begin{eqnarray}
a_{11} & = &  4(I_f^{3L} \cos^2 \theta_{\tilde f}  - \sin^2 \theta_W e_f) \ , 
\nonumber \\
a_{22} & = &  4(I_f^{3L} \sin^2 \theta_{\tilde f} - \sin^2 \theta_W e_f) \ , 
\nonumber \\
a_{12} & = &  a_{21}= -2 I_f^{3L} \sin 2 \theta_{\tilde f} \, ,
\label{acoup}
\end{eqnarray}
where $I_f^{3L}$ is the third component of the weak isospin of the fermion $f$.
$G^{h^0}_{ij}$, $G^{H^0}_{ij}$, and $G^{A^0}_{ij}$ are the couplings of the
sfermions $\tilde f_i$ and $\tilde f_j$ (i, j =1, 2) to the light, the heavy, 
and the pseudoscalar Higgs boson, respectively. These couplings depend on
$A_f$, $\mu$, $\tan \beta$, $\cos \theta_{\tilde f}$, $\cos \alpha$, where 
$\alpha$ is the mixing angle of $h^0$ and $H^0$. The explicit
form of these couplings is given in the Appendix.

\section{Numerical Results}

In this Section we present the numerical results for sfermion production in
the various channels. In this analysis we have taken: $\alpha(m_Z) = 1/129$,
$\sin^2 \theta_W = 0.23$, $m_Z = 91.187$~GeV, $m_t = 175$~GeV, $m_b = 5$~GeV,
and $m_\mu = 105.66$~MeV. For the calculation of the cross section we 
need the total decay widths of the Higgs bosons, where we calculate all
possible two--body decay modes that are allowed at tree--level
\cite{bartl96b}. For this 
calculation we fix $M = 120$~GeV and use the GUT relation 
$M' = 5/3 \tan^2 \theta_W  M$, where $M$ and $M'$ are the $SU(2)$ and $U(1)$
gaugino masses, respectively.
As usual we have included radiative corrections in the calculation of 
$m_{h^0}$, $m_{H^0}$, and $\cos \alpha$ using \cite{ellis}.

\begin{figure}[t!] 
\begin{picture}(220,500)
\put(-30,-55){\mbox{\psfig{file=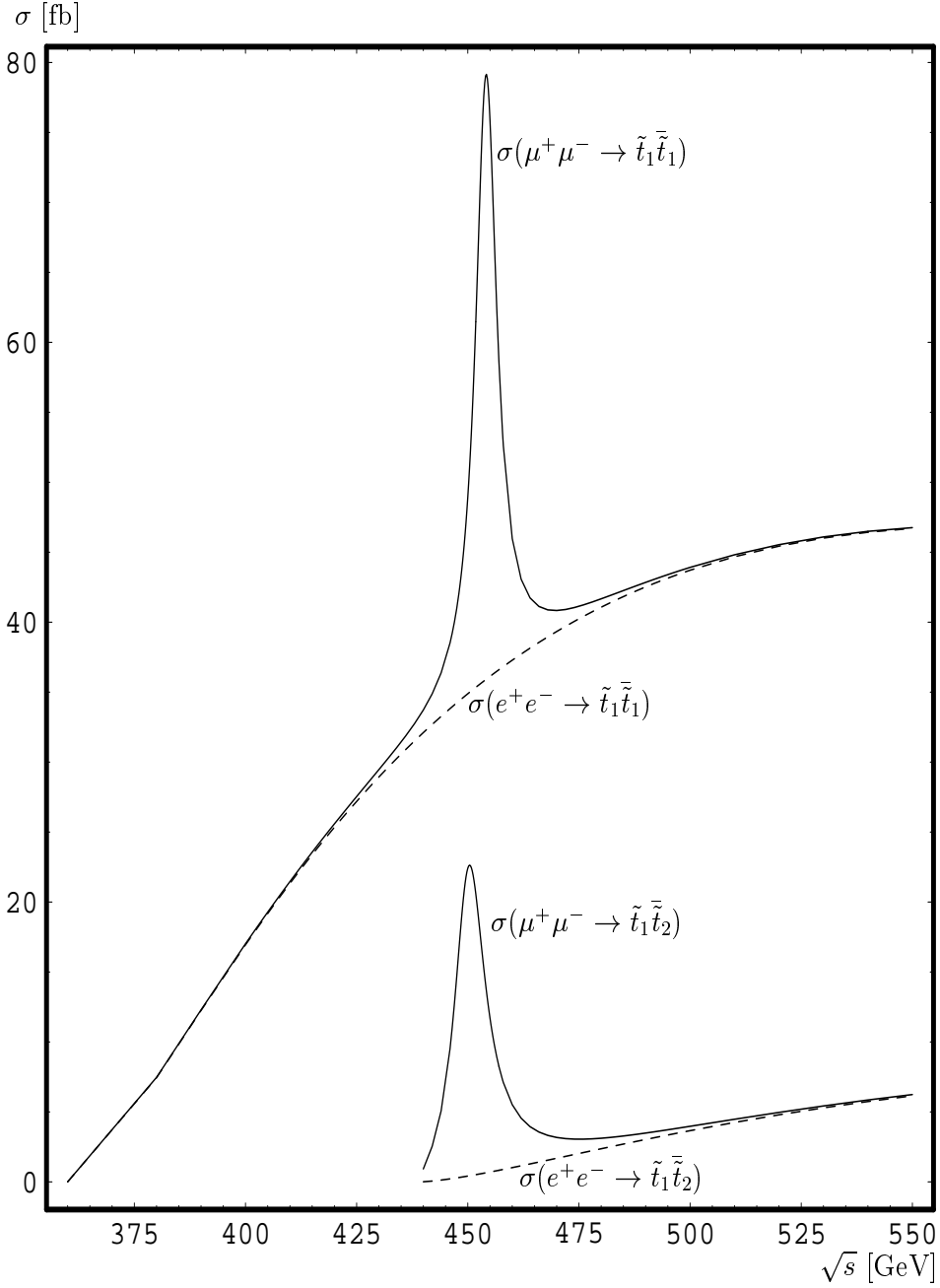,height=8.8in,width=6.5in}}}
\end{picture}
\vspace{10pt}
\caption{Cross sections for stop production as a function of $\sqrt{s}$
for $m_{\tilde t_1} = 180$~GeV, $m_{\tilde t_2} = 260$~GeV, 
$\cos \theta_{\tilde t} = -0.556$,
$m_{\tilde b_1} = 175$~GeV, $m_{\tilde b_2} = 195$~GeV, 
$\cos \theta_{\tilde b} = 0.9$, $\mu = 300$~GeV,
$M = 120$~GeV, $\tan \beta = 3$, and $m_{A^0} = 450$~GeV. The full lines 
show the cross sections at a $\mu^+ \mu^-$ collider and the dashed lines 
the corresponding ones at an $e^+ e^-$ collider.}
\label{stopsqrts}
\end{figure}

In Fig.~\ref{stopsqrts} we show the total cross sections for stop
production as a function of $\sqrt{s}$ for $\mu = 300$~GeV,
$\tan \beta = 3$,
$m_{\tilde t_1} = 180$~GeV, $m_{\tilde t_2} = 260$~GeV, 
$\cos \theta_{\tilde t} = -0.556$, and $m_{A^0} = 450$~GeV. For the total
widths of the Higgs bosons we take
$m_{\tilde b_1} = 175$~GeV, $m_{\tilde b_2} = 195$~GeV, 
$\cos \theta_{\tilde b} = 0.9$, $M_{\tilde L} = 170$~GeV, 
$M_{\tilde E} = 150$~GeV, and $A_\tau = 300$~GeV. The full lines 
show the total cross sections and the dashed lines the gauge boson 
contributions.
The latter ones are identical with the cross sections of
$e^+ e^- \to \tilde t_i \bar{\tilde t_j}$. 
For $\tilde t_1 \bar{\tilde t_1}$ production
the peak results from the $H^0$ exchange leading to an enhancement of
$\sim 40$~fb compared to the gauge boson contribution. For 
$\tilde t_1 \bar{\tilde t_2}$ production the peak is an overlap of the $H^0$ 
and $A^0$ resonances because $m_{A^0} \simeq m_{H^0}$ and 
the widths of $A^0$ and $H^0$ are of the order of several GeV
(see e.g. \cite{higgshunter,bartl97a,djouadi97}). Note that the Higgs boson
contribution is much larger than the gauge boson contribution.
We have found that the forward--backward asymmetry $A_{FB}$ is $\sim 10^{-4}$
at its maximum. Therefore, a rather high luminosity
would be needed to measure it.

\begin{figure}[t!] 
\begin{picture}(220,500)
\put(-30,-55){\mbox{\psfig{file=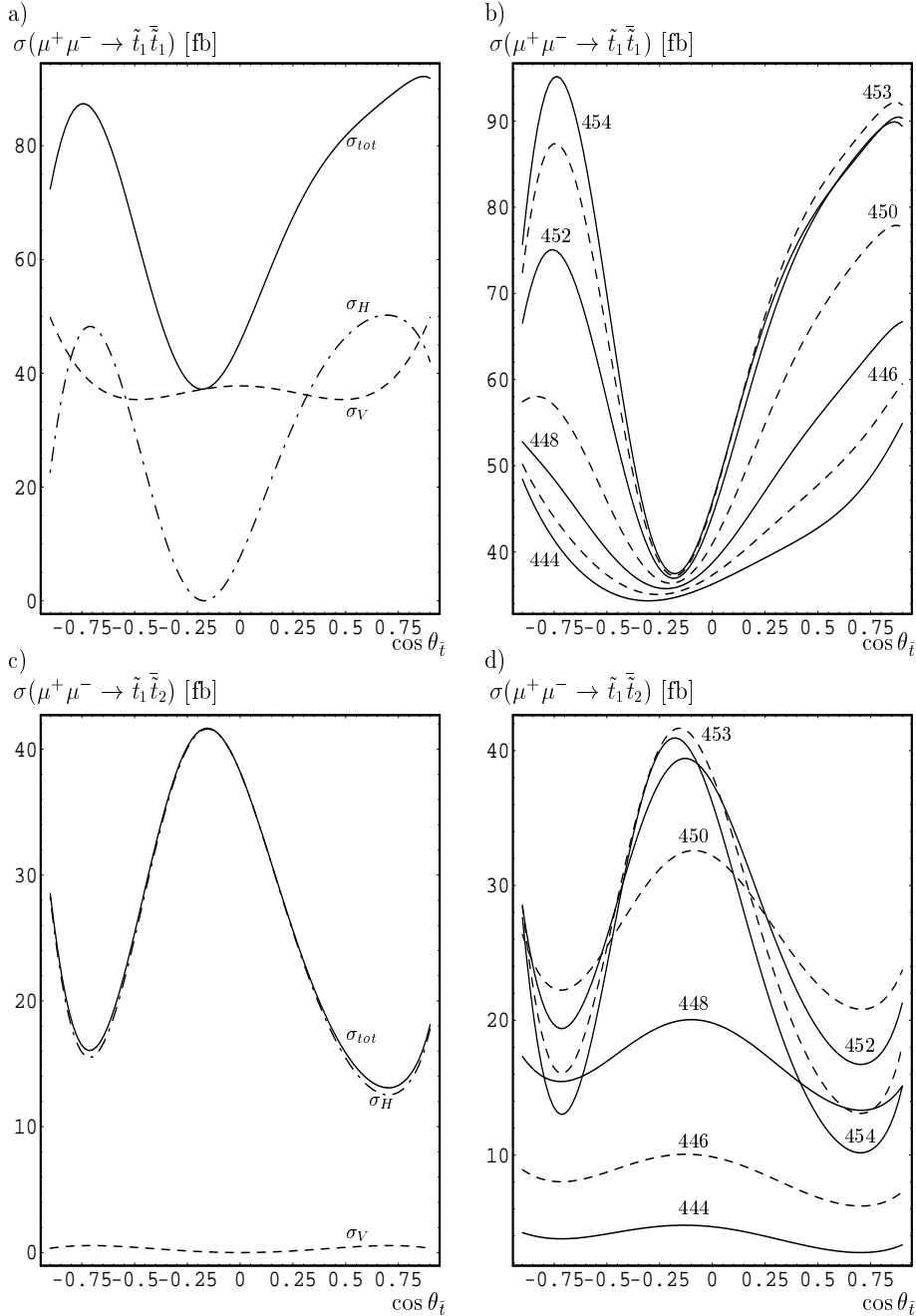,height=8.8in,width=6.5in}}}
\end{picture}
\vspace{10pt}
\caption{Cross sections for stop production as a function of 
$\cos \theta_{\tilde t}$ for $m_{\tilde t_1} = 180$~GeV, 
$m_{\tilde t_2} = 260$~GeV, $A_b = 300$~GeV, $\mu = 300$~GeV,
$M = 120$~GeV, $\tan \beta = 3$, and $m_{A^0} = 450$~GeV. 
In a) and c) $\sqrt{s} = 453$~GeV and the 
graphs correspond to: total cross section $\sigma_{tot}$ (full line), Higgs
boson contribution $\sigma_H$ (dashed-dotted line), and gauge boson 
contribution $\sigma_V$ (dashed line). In b) and d) the cross section is shown
for various $\sqrt{s}$ values (in GeV): 444, 448, 452, 454 (full lines) and
446, 450, 453 (dashed lines).} 
\label{stopcos}
\end{figure}

In Fig.~\ref{stopcos} we show the production cross sections for 
$\mu^+ \mu^- \to \tilde t_1 \bar{\tilde t}_1$ and  
$\mu^+ \mu^- \to \tilde t_1 \bar{\tilde t}_2$ (without including the charge
conjugate state) as a function of 
$\cos \theta_{\tilde t}$. We have
chosen the following procedure for the calculation of the parameters: We fix 
$m_{\tilde t_1} = 180$~GeV, $m_{\tilde t_2} = 260$~GeV, $A_b = 300$~GeV, and
$\mu$, $\tan \beta$, $m_{A^0}$, $M_{\tilde L}$, $M_{\tilde E}$, 
$A_\tau$ as above. We calculate $M_{\tilde Q}$,
$M_{\tilde U}$, and $A_t$ from the stop masses and mixing angle. We take
$M_{\tilde D} = 1.12 \, M_{\tilde Q}$ (we have checked, that our results are 
not sensitive to this assumption). These parameters
are then used for the calculation of $m_{\tilde b_1}$, $m_{\tilde b_2}$,
$\cos \theta_{\tilde b}$, $m_{h^0}$, $m_{H^0}$, $\cos \alpha$, 
$\Gamma_{H^0}$, and $\Gamma_{A^0}$. We have chosen
this procedure to minimize the dependence of the physical Higgs quantities
on $\cos \theta_{\tilde t}$.
In Fig.~\ref{stopcos}a we show the
total cross section $\sigma(\mu^+ \mu^- \to \tilde t_1 \bar{\tilde t}_1)$, the
Higgs boson contribution, and the gauge boson contribution for 
$\sqrt{s} = 453$~GeV. The Higgs boson contribution depends on the
sign of $\cos \theta_{\tilde t}$. This leads to a dependence of the total
cross section on the sign of $\cos \theta_{\tilde t}$ contrary to the
case of an $e^+ e^-$ collider, where the cross section depends only on
$\cos^2 \theta_{\tilde t}$ \cite{Bartl97}. Note that the Higgs boson
contribution can be larger than the gauge boson contribution. This is in
particular the case for large mixing in the stop sector because the Higgs
boson couples more strongly to the left--right combination of the stops than to
the left--left or right--right combinations. The Higgs boson contribution 
vanishes for
$\cos \theta_{\tilde t} \simeq -0.25$ because the corresponding coupling is
zero for the parameters chosen.  
One can disentangle the Higgs boson contribution from the gauge boson
part by measuring the differential cross section. As can be seen in 
Eq.~(\ref{sigdiff}) the
Higgs boson contribution of the differential cross section does not depend
on $\vartheta$ whereas the gauge boson contribution shows a 
$\sin^2 \vartheta$ shape. We can safely neglect the
interference terms between gauge and Higgs bosons because they are rather small
as we have seen above. In Fig.~\ref{stopcos}b the total cross
section is shown for various values of $\sqrt{s}$ between 444~GeV and 454~GeV.
For larger values of $\sqrt{s}$ the cross section is decreasing because  
$m_{H^0}$ varies between 452.8~GeV ($\cos \theta_{\tilde t} \sim 0.71$) and
454.2~GeV ($\cos \theta_{\tilde t} \sim -0.71$). The $\cos \theta_{\tilde t}$ 
dependence of $m_{H^0}$ is also the reason for 
$\sigma(\sqrt{s}=454) > (<) \sigma(\sqrt{s}=453)$ if 
$\cos \theta_{\tilde t} < (>) -0.25$.

In Fig.~\ref{stopcos}c we show 
$\sigma(\mu^+ \mu^- \to \tilde t_1 \bar{\tilde t}_2)$, 
the Higgs boson contribution and
the gauge boson contribution as a function of $\cos \theta_{\tilde t}$. 
An interesting feature here is that for 
$\cos \theta_{\tilde t} = 0$ only the Higgs bosons contribute. Moreover, the
contribution of the Higgs bosons $H_0$ and $A_0$ is generally larger than 
that of the gauge boson. The total
cross section again depends on the sign of $\cos \theta_{\tilde t}$. 
In Fig.~\ref{stopcos}d we show the total cross section for various 
values of $\sqrt{s}$.
The shift of the peak is due the dependence of $m_{H^0}$ and $\cos \alpha$
on $A_t$ which is calculated from $\cos \theta_{\tilde t}$.
Note that the coupling $H^0 \tilde t_1 \tilde t_2$ is large
in the range where the coupling $H^0 \tilde t_1 \tilde t_1$ is small and vice
versa. The minima near $|\cos \theta_{\tilde t}| \simeq 0.71$ are due to 
the vanishing of the coupling $H^0 \tilde t_1 \tilde t_2$.

\begin{figure}[t!] 
\begin{picture}(220,500)
\put(-30,-55){\mbox{\psfig{file=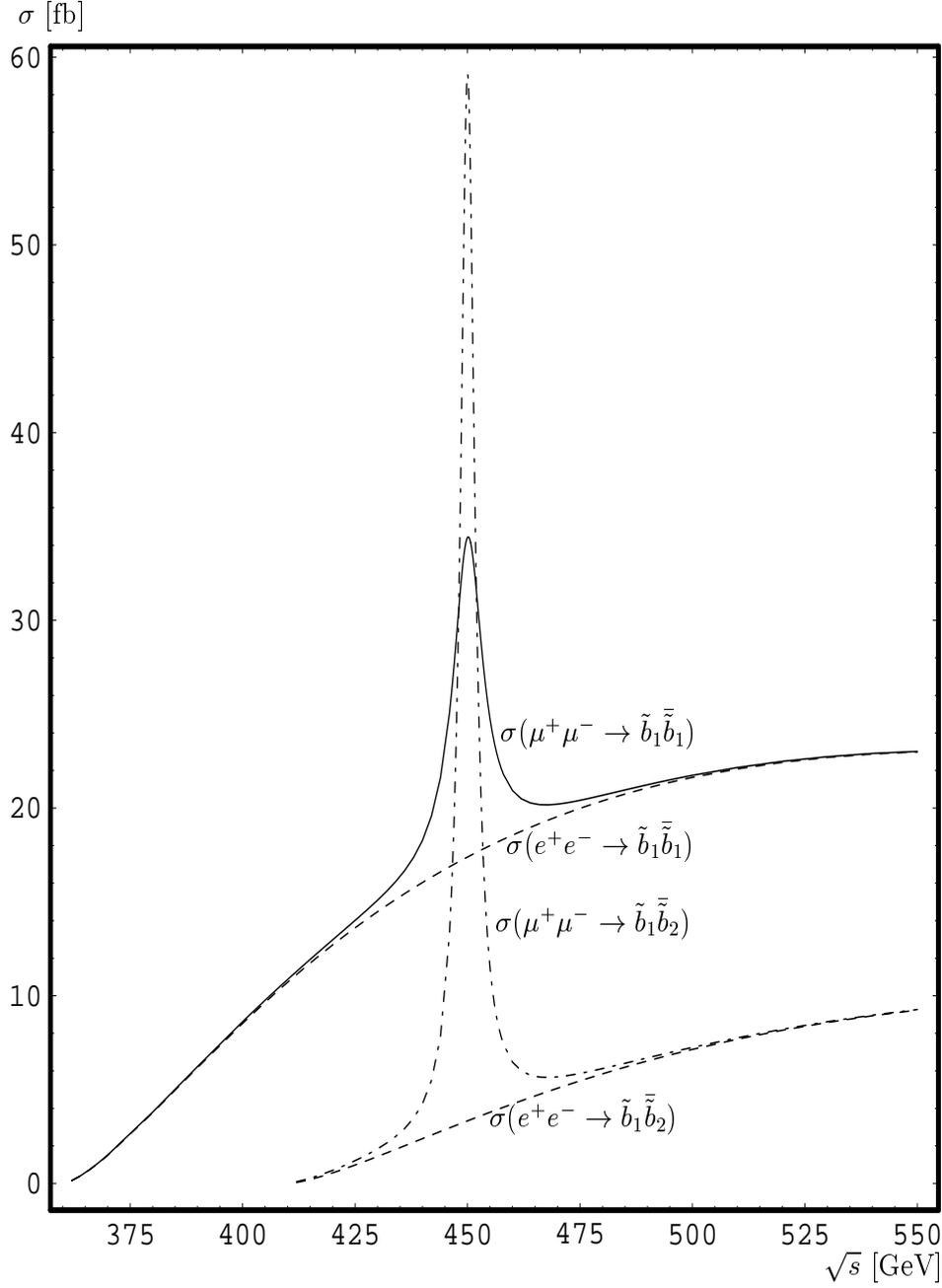,height=8.8in,width=6.5in}}}
\end{picture}
\vspace{10pt}
\caption{Cross sections for sbottom production as a function of $\sqrt{s}$ for
$m_{\tilde b_1} = 180$~GeV, $m_{\tilde b_2} = 230$~GeV, 
$\cos \theta_{\tilde b} = 0.755$, $m_{\tilde t_1} = 160$~GeV, 
$m_{\tilde t_2} = 300$~GeV, $\cos \theta_{\tilde t} = 0.615$, $\mu = 291$~GeV,
$M = 120$~GeV, $\tan \beta = 8$, and $m_{A^0} = 450$~GeV.}
\label{sbotsqrts}
\end{figure}

In Fig.~\ref{sbotsqrts} the cross sections for sbottom production are shown as
a function of $\sqrt{s}$ for   
$m_{\tilde b_1} = 180$~GeV, $m_{\tilde b_2} = 230$~GeV, 
$\cos \theta_{\tilde b} = 0.755$, $m_{\tilde t_1} = 160$~GeV, 
$m_{\tilde t_2} = 300$~GeV, $\cos \theta_{\tilde t} = 0.615$, $\mu = 291$~GeV,
$\tan \beta = 8$, and $m_{A^0} = 450$~GeV. $M_{\tilde L}$, $M_{\tilde E}$, 
and $A_\tau$ are taken as above. It is interesting 
that $\sigma(\mu^+ \mu^- \to \tilde b_1 \bar{\tilde b}_2)$ is 
$\sim 20$ times larger than $\sigma(e^+ e^- \to \tilde b_1 \bar{\tilde b}_2)$ 
at $\sqrt{s} = m_{A^0}$. This has two implications: First, one gets a cross 
section that is large enough to be measured even with an integrated luminosity 
of 10~fb$^{-1}$. Second, the $b_1 \bar{\tilde b}_2$ cross section is even 
larger than the $\tilde b_1 \bar{\tilde b}_1$ production cross section.
Note that we only show the cross section for $b_1 \bar{\tilde b}_2$ whereas in
the experiment one can measure the cross section of 
$\tilde b_1 \bar{\tilde b}_2 + \bar{\tilde b}_1 \tilde b_2$. 

\begin{figure}[t!] 
\begin{picture}(220,500)
\put(-30,-55){\mbox{\psfig{file=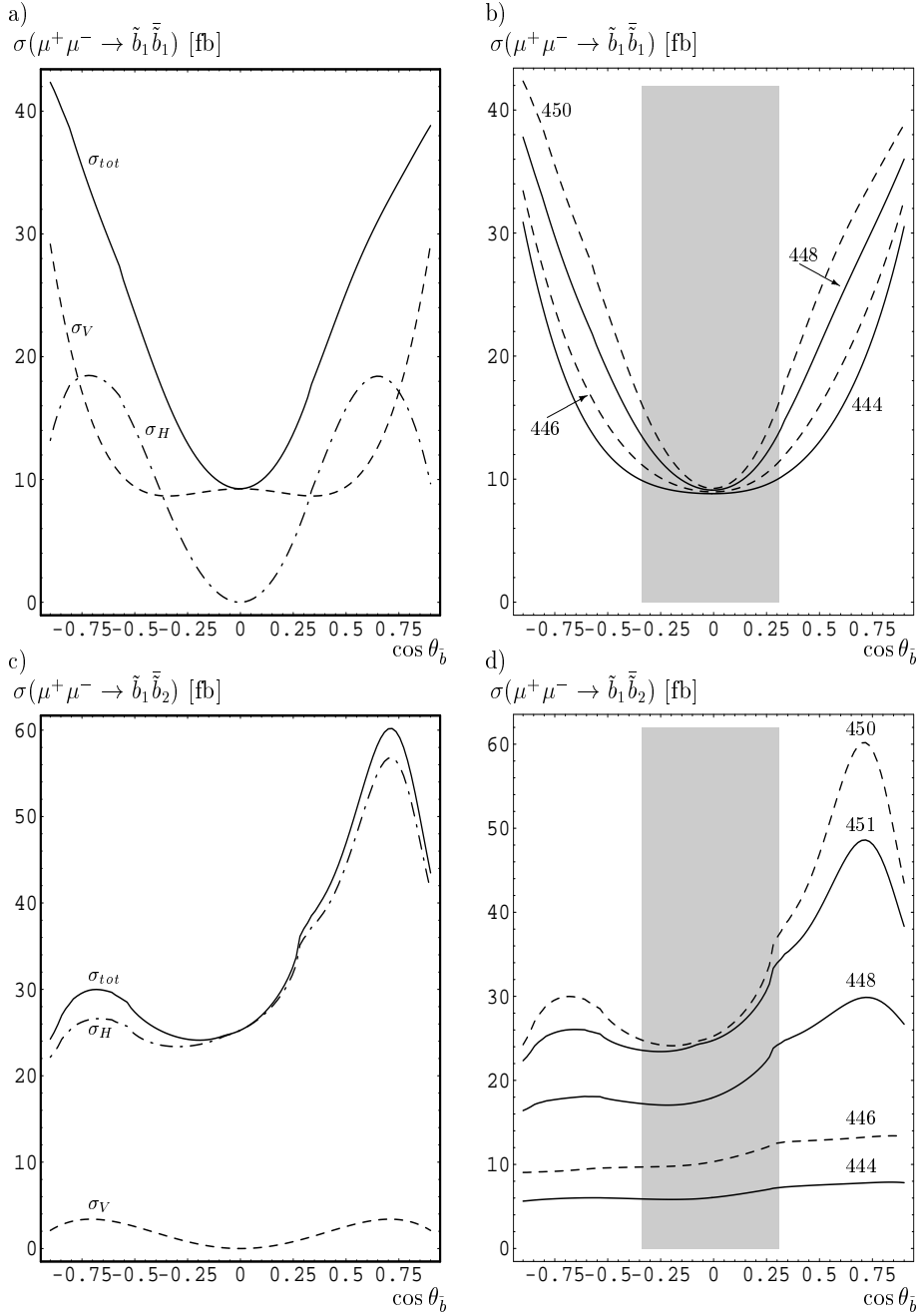,height=8.8in,width=6.5in}}}
\end{picture}
\vspace{10pt}
\caption{Cross sections for sbottom production as a function of 
$\cos \theta_{\tilde b}$ for $m_{\tilde b_1} = 180$~GeV, 
$m_{\tilde b_2} = 230$~GeV, $A_b = 300$~GeV,  $m_{\tilde t_1} = 160$~GeV, 
$m_{\tilde t_2} = 300$~GeV,
$M = 120$~GeV, $\tan \beta = 8$, and $m_{A^0} = 450$~GeV. 
In a) and c) $\sqrt{s} = 450$~GeV and the 
graphs correspond to: total cross section $\sigma_{tot}$ (full line), Higgs boson
boson contribution $\sigma_H$ (dashed-dotted line), and gauge boson 
contribution $\sigma_V$ (dashed line). In b) and d) the cross section is shown
for various $\sqrt{s}$ values (in GeV): 444, 448, 451 (full lines) and
446, 450 (dashed lines). The gray area is excluded by LEP2 
($m_{\tilde \chi^+_1} < 90$~GeV).} 
\label{sbotcos}
\end{figure}

In Fig.~\ref{sbotcos} we show the cross sections for sbottom production as a 
function of  $\cos \theta_{\tilde b}$ for $m_{\tilde b_1} = 180$~GeV, 
$m_{\tilde b_2} = 230$~GeV, $A_b = 300$~GeV, $m_{\tilde t_1} = 160$~GeV, 
$m_{\tilde t_2} = 300$~GeV, $\tan \beta = 8$, and $m_{A^0} = 450$~GeV. 
$M_{\tilde L}$, $M_{\tilde E}$, and 
$A_\tau$ are taken as above. We have calculated
the other parameters in the following way: From  $m_{\tilde b_1}$, 
$m_{\tilde b_2}$, $\cos \theta_{\tilde b}$, and $A_b$ we get
$M_{\tilde Q}$, $M_{\tilde D}$, and $\mu$. We then take $M_{\tilde Q}$,
$m_{\tilde t_1}$, and $m_{\tilde t_2}$ to calculate $M_{\tilde U}$,
$\cos \theta_{\tilde t}$, and $A_t$. 
There are similarities to the stop case: For $\tilde b_1 \bar{\tilde b}_1$ 
production the Higgs boson contribution can be larger than the gauge boson
part (Fig.~\ref{sbotcos}a). In the case of $\tilde b_1 \bar{\tilde b}_2$ 
production, the
contribution of  the Higgs bosons is much larger than that of
the gauge boson
(Fig.~\ref{sbotcos}c). The main difference compared to the stop case is that
the asymmetry in the sign of $\cos \theta_{\tilde b}$ is much more pronounced
in the $\tilde b_1 \bar{\tilde b}_2$ channel than in the 
$\tilde b_1 \bar{\tilde b}_1$
channel as a consequence of the corresponding couplings
(Fig.~\ref{sbotcos}b and d). The peak at $\cos \theta_{\tilde b} \simeq 0.71$
in Fig.~\ref{sbotcos}d
results not only from the couplings but also from the fact that the total
decay width of $A^0$ has a minimum there. 

\begin{figure}[t!] 
\begin{picture}(220,500)
\put(-30,-55){\mbox{\psfig{file=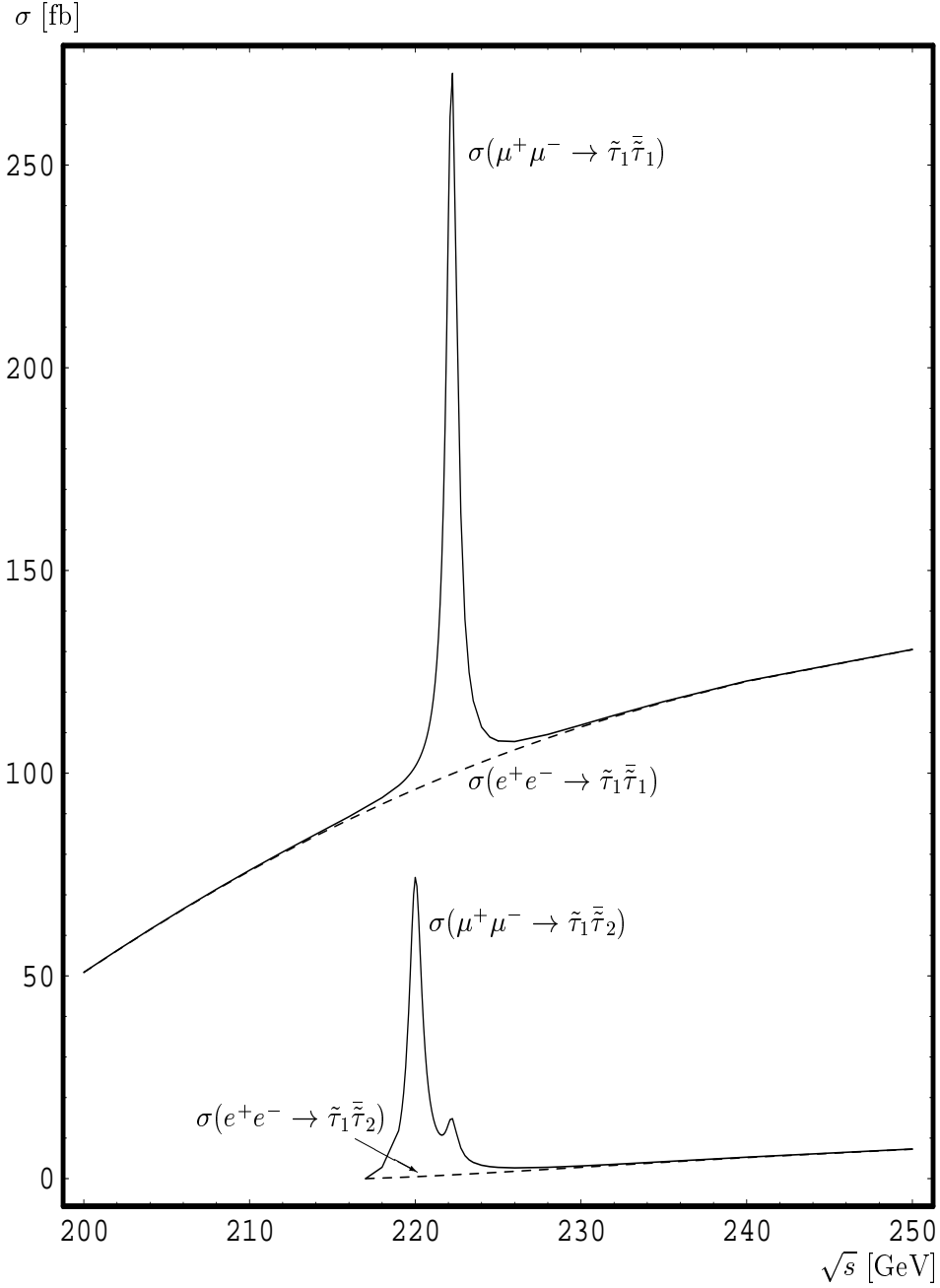,height=8.8in,width=6.5in}}}
\end{picture}
\vspace{10pt}
\caption{Cross sections for stau production as a function of $\sqrt{s}$ for
$m_{\tilde \tau_1} = 90$~GeV,
$m_{\tilde \tau_2} = 127$ GeV, $\cos \theta_{\tilde \tau} = 0.594$,
$M_{\tilde Q} = 300$~GeV, $M_{\tilde U} = 270$~GeV,
$M_{\tilde D} = 330$~GeV, $A_t = 350$~GeV, $A_b = 350$~GeV, 
$\mu = 300$~GeV,
$M = 120$~GeV, $\tan \beta = 8$, and $m_{A^0} = 220$~GeV. The full lines 
show the cross sections at a $\mu^+ \mu^-$ collider and the dashed lines 
the corresponding ones at an $e^+ e^-$ collider.}
\label{stausqrts}
\end{figure}

In Fig.~\ref{stausqrts} the stau production cross sections
are shown as a function of $\sqrt{s}$ for $m_{\tilde \tau_1} = 90$~GeV,
$m_{\tilde \tau_2} = 127$~GeV, $\cos \theta_{\tilde \tau} = 0.594$,
$M_{\tilde Q} = 300$~GeV, $M_{\tilde U} = 270$~GeV,
$M_{\tilde D} = 330$~GeV, $A_t = A_b = 350$~GeV, $\mu = 300$~GeV, 
$\tan \beta = 8$, and $m_{A^0} = 220$~GeV. The parameters are
chosen such that the Higgs boson cannot decay into squarks or the top quark.
Therefore, the total decay widths of $H^0$ and $A^0$ are one order of magnitude
smaller than in the previous examples.
This leads to the large peaks at $\sqrt{s} = m_{A^0}, m_{H^0}$. Moreover, the
decay widths are so small that one can see two peaks in the case of 
$\tilde \tau_1 \bar{\tilde \tau_2}$ production. It should be possible to
observe both peaks in a real experiment because of the good energy resolution 
of a $\mu^+ \mu^-$ collider \cite{haber}.

\begin{figure}[t!] 
\begin{picture}(220,500)
\put(-30,-55){\mbox{\psfig{file=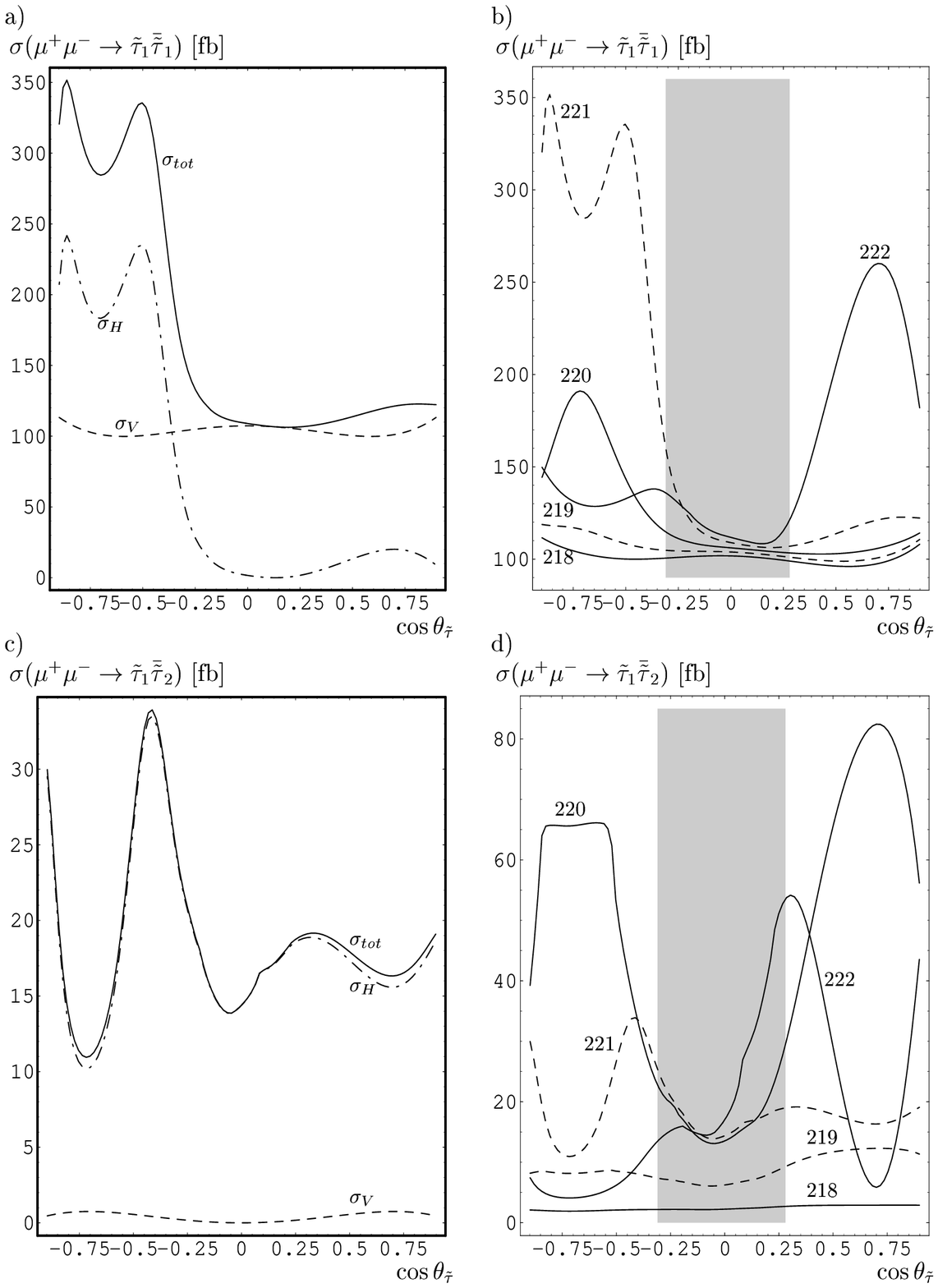,height=8.8in,width=6.5in}}}
\end{picture}
\vspace{10pt}
\caption{Cross sections for $\tilde \tau$ production as a function of 
$\cos \theta_{\tilde \tau}$ for $m_{\tilde \tau_1} = 90$~GeV,
$m_{\tilde \tau_2} = 127$ GeV, $A_\tau = 300$~GeV,
$M = 120$~GeV, $\tan \beta = 8$, and $m_{A^0} = 220$~GeV.  In a) and c) 
$\sqrt{s} = 221$~GeV and the 
graphs correspond to: total cross section $\sigma_{tot}$ (full line), Higgs
boson contribution $\sigma_H$ (dashed-dotted line), and gauge boson 
contribution $\sigma_V$ (dashed line). In b) and d) the cross section is shown
for various $\sqrt{s}$ values (in GeV): 218, 220, 222 (full lines) and
219, 221 (dashed lines).  The gray area is excluded by LEP2 
($m_{\tilde \chi^+_1} < 90$~GeV).} 
\label{staucos}
\end{figure}

\begin{figure}[t!] 
\begin{picture}(220,500)
\put(-30,-55){\mbox{\psfig{file=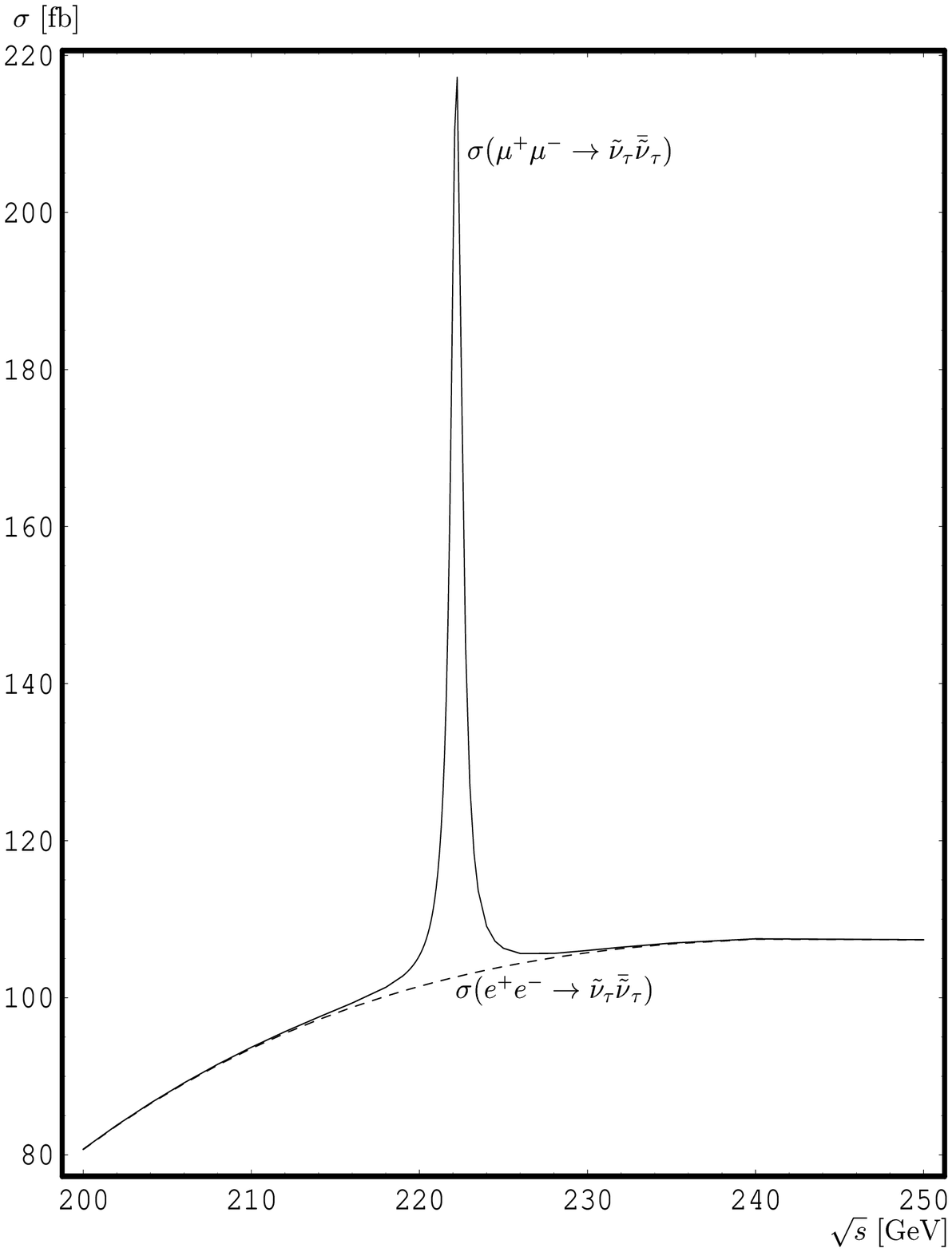,height=8.8in,width=6.5in}}}
\end{picture}
\vspace{10pt}
\caption{Cross sections for sneutrino production as a function of $\sqrt{s}$
for $m_{\tilde \nu_\tau} = 83.6$~GeV and the other parameters as in 
Fig.~\ref{stausqrts}.  The full line 
shows the cross sections at a $\mu^+ \mu^-$ collider and the dashed line 
the corresponding one at an $e^+ e^-$ collider.}
\label{sneutsqrts}
\end{figure}

In Fig.~\ref{staucos} the cross sections for stau production are presented as
a function of $\cos \theta_{\tilde \tau}$ for $m_{\tilde \tau_1} = 90$~GeV,
$m_{\tilde \tau_2} = 127$~GeV, $A_\tau = 300$~GeV. $\tan \beta$, 
$m_{A^0}$, and the squark parameters are taken as above. We have calculated 
$M_{\tilde L}$, $M_{\tilde E}$ and $\mu$ from $m_{\tilde \tau_{1,2}}$ and
$\cos \theta_{\tilde \tau}$. With these and the other parameters we have 
calculated 
$m_{\tilde \nu_\tau}$, $m_{H^0}$, $\cos \alpha$, $\Gamma_{H^0}$, and
$\Gamma_{A^0}$. Note that the masses, mixing angle, and decay widths of the
Higgs bosons depend indirectly on 
$\cos \theta_{\tilde \tau}$ due to the induced change in $\mu$. This fact
leads to the observed shifts of the maximal cross section with $\sqrt{s}$ in
Fig.~\ref{staucos}b and d. In $\tilde \tau_1 \bar{\tilde \tau_2}$ 
production the Higgs boson contribution is much larger than the gauge boson 
contribution (Fig.\ref{staucos}c) similar
to the squark production. This is particularly important, because in this case
the production cross section is most likely too small to be seen at an
$e^+ e^-$ collider \cite{Bartl97,desy}. 
The cross sections depend strongly on the sign of 
$\cos \theta_{\tilde \tau}$ in both channels (Fig.~\ref{staucos}a and c).
Note also that, for the parameters chosen in the Higgs couplings to the
staus, the gauge couplings are of the same size as the Yukawa couplings
(Eqs.(\ref{higgsstaua})--(\ref{higgsstaub})):

In Fig.~\ref{sneutsqrts} we show the total cross sections for sneutrino 
production as a function of $\sqrt{s}$ for the same parameters as in
Fig.~\ref{stausqrts} (implying $m_{\tilde \nu_\tau} = 83.6$~GeV).
The large peak at $\sqrt{s} = m_{H^0}$ is due to the small total decay width 
of $H^0$ ($\Gamma_{H^0} \simeq 0.8$~GeV) and due to the coupling 
$H^0 \tilde \nu_\tau \tilde \nu_\tau$ which is a gauge coupling 
(see Eq.~(\ref{coupsneut})\,) stemming from a D--term.

\section{Summary}

We have studied sfermion pair production 
in $\mu^+ \mu^-$ annihilation focusing on the impact of the Higgs boson
resonances in these processes. We have seen that the production cross sections 
can be considerably enhanced at these resonances for all sfermions of the third
generation. The most important results are:
First, the production cross sections depend on the sign of 
$\cos \theta_{\tilde f}$.
Second, the Higgs boson contributions dominate the production cross section
of $\tilde f_1 \bar{\tilde f}_2$. We have seen that the cross
sections can be large enough to be studied at a $\mu^+ \mu^-$ collider even if
the corresponding cross sections are too small to be measured at an $e^+ e^-$ 
collider.
Third, the Higgs boson contribution can even be larger than the gauge boson 
contributions in the $\tilde f_1 \bar{\tilde f_1}$ channel. 
From these facts we conclude that a $\mu^+ \mu^-$ collider is an excellent
machine for obtaining
important information on the $H^0 \tilde f_i \tilde f_j$ and 
$A^0 \tilde f_1 \tilde f_2$ couplings. 

\subsection*{Acknowledgments}

We are very grateful to M.~Carena and S.~Protopopescu for their kind invitation
to the ''Workshop on Physics at the
First Muon Collider and the Front End of a Muon Collider''
(FNAL, Batavia, Illinois, USA, November 6 -- 9, 1997) where this study was 
initiated.
This work was supported by the "Fonds zur F\"orderung der
wissenschaftlichen Forschung" of Austria, project no. P10843--PHY.

\begin{appendix}
\section{Higgs couplings}

In this section we list the couplings of the neutral Higgs bosons to
sfermions. We concentrate here on the couplings of $H^0$ and $A^0$
which are important for our investigation. One can
get the couplings of $h^0$ from those of $H^0$ by the replacements:
$\cos \alpha \to -\sin \alpha$ and $\sin \alpha \to \cos \alpha$.
\\ Higgs couplings to stops:
\begin{eqnarray}
G^{H^0}_{11} & = &   
- \frac{m_Z \cos ( \alpha + \beta)}{4 \, \cos \theta_W}
- \frac{m_Z}{2 \, \cos \theta_W} 
  \left(\frac{1}{2} - \frac{4}{3} \sin^2 \theta_W \right) 
       \cos (\alpha + \beta) \cos 2 \theta_{\tilde t} \nonumber \\ 
& &
-\frac{m^2_t \sin \alpha}{m_W \sin \beta}
+ \frac{m_t}{2 \, m_W \sin \beta} \left(\mu \cos \alpha 
             - A_t \sin \alpha \right) \sin 2 \theta_{\tilde t}, \\ 
G^{H^0}_{12} & = &   
\frac{m_Z}{2 \, \cos \theta_W} 
\left(\frac{1}{2} - \frac{4}{3} \sin^2 \theta_W \right) 
       \cos (\alpha + \beta) \sin 2 \theta_{\tilde t} \nonumber \\
& & + \frac{m_t}{2 \, m_W \sin \beta} 
  \left(\mu \cos \alpha - A_t \sin \alpha \right) \cos 2 \theta_{\tilde t}, \\ 
G^{H^0}_{22} & = &   
- \frac{m_Z \cos ( \alpha + \beta)}{4 \, \cos \theta_W}
+ \frac{m_Z}{2 \, \cos \theta_W} 
  \left(\frac{1}{2} - \frac{4}{3} \sin^2 \theta_W \right) 
       \cos (\alpha + \beta) \cos 2 \theta_{\tilde t} \nonumber \\
& & -\frac{m^2_t \sin \alpha}{m_W \sin \beta} 
- \frac{m_t}{2 \, m_W \sin \beta} 
  \left(\mu \cos \alpha - A_t \sin \alpha \right) \sin 2 \theta_{\tilde t}, \\ 
G^{A^0}_{12} & = & -G^{A^0}_{21} =
 \frac{m_t}{2 \, m_W} (A_t \cot \beta + \mu).
\end{eqnarray}
Higgs couplings to sbottoms:
\begin{eqnarray}
G^{H^0}_{11} & = &  
- \frac{m_Z \cos ( \alpha + \beta)}{4 \, \cos \theta_W}
- \frac{m_Z}{2 \, \cos \theta_W} 
  \left(-\frac{1}{2} + \frac{2}{3} \sin^2 \theta_W \right) 
       \cos (\alpha + \beta) \cos 2 \theta_{\tilde b} \nonumber \\
& & -\frac{m^2_b \cos \alpha}{m_W \cos \beta} 
+ \frac{m_b}{2 \, m_W \cos \beta} 
  \left(\mu \sin \alpha - A_b \cos \alpha \right) \sin 2 \theta_{\tilde b}, \\ 
G^{H^0}_{12} & = & 
\frac{m_Z}{2 \, \cos \theta_W} 
\left(-\frac{1}{2} + \frac{2}{3} \sin^2 \theta_W \right)
       \cos (\alpha + \beta) \sin 2 \theta_{\tilde b} \nonumber \\ 
& & + \frac{m_b}{2 \, m_W \cos \beta} 
 \left(\mu \sin \alpha - A_b \cos \alpha \right) \cos 2 \theta_{\tilde b}, \\ 
G^{H^0}_{22} & = &  
- \frac{m_Z \cos ( \alpha + \beta)}{4 \, \cos \theta_W}
+ \frac{m_Z}{2 \, \cos \theta_W} 
  \left(-\frac{1}{2} + \frac{2}{3} \sin^2 \theta_W \right) 
       \cos (\alpha + \beta) \cos 2 \theta_{\tilde b} \nonumber \\ 
& & -\frac{m^2_b \cos \alpha}{m_W \cos \beta} 
- \frac{m_b}{2 \, m_W \cos \beta}
  \left(\mu \sin \alpha - A_b \cos \alpha \right) \sin 2 \theta_{\tilde b}, \\
G^{A^0}_{12} & = & -G^{A^0}_{21} = \frac{m_b}{2 \, m_W} (A_b \tan \beta + \mu).
\end{eqnarray}
Higgs coupling to the tau--sneutrino:
\begin{eqnarray}
G^{H^0}_{11} & = &  
-\frac{m_Z \cos ( \alpha + \beta)}{2 \, \cos \theta_W}. 
\label{coupsneut}
\end{eqnarray} 
Higgs couplings to staus:
\begin{eqnarray}
G^{H^0}_{11} & = & 
- \frac{m_Z \cos ( \alpha + \beta)}{4 \, \cos \theta_W} 
- \frac{m_Z}{2 \, \cos \theta_W} \left(-\frac{1}{2} + \sin^2 \theta_W \right) 
       \cos (\alpha + \beta) \cos 2 \theta_{\tilde \tau} \nonumber \\ 
& & -\frac{m^2_\tau \cos \alpha}{m_W \cos \beta} 
+ \frac{m_\tau}{2 \, m_W \cos \beta} 
  \left(\mu \sin \alpha - A_\tau \cos \alpha \right) 
    \sin 2 \theta_{\tilde \tau},
\label{higgsstaua}  \\
G^{H^0}_{12} & = & 
\frac{m_Z}{2 \, \cos \theta_W} \left(-\frac{1}{2} + \sin^2 \theta_W \right)
       \cos (\alpha + \beta) \sin 2 \theta_{\tilde \tau} \nonumber \\
& & + \frac{m_\tau}{2 \, m_W \cos \beta} 
     \left(\mu \sin \alpha - A_\tau \cos \alpha \right)
     \cos 2 \theta_{\tilde \tau}, \\
G^{H^0}_{22} & = & 
- \frac{m_Z \cos ( \alpha + \beta)}{4 \, \cos \theta_W} 
+ \frac{m_Z}{2 \, \cos \theta_W} \left(-\frac{1}{2} + \sin^2 \theta_W \right) 
       \cos (\alpha + \beta) \cos 2 \theta_{\tilde \tau} \nonumber \\
& & -\frac{m^2_\tau \cos \alpha}{m_W \cos \beta} 
- \frac{m_\tau}{2 \, m_W \cos \beta}
  \left(\mu \sin \alpha - A_\tau \cos \alpha \right)
   \sin 2 \theta_{\tilde \tau},
\label{higgsstaub} \\ 
G^{A^0}_{12} & = & - G^{A^0}_{21} =  \frac{m_\tau}{2 \, m_W}
 (A_\tau \tan \beta + \mu).
\end{eqnarray}
\end{appendix}

\newpage

\newpage

\newpage

\newpage

\newpage

\newpage

\newpage

\begin{thebibliography}{99}
\bibitem{susy} H.~P.~Nilles, Phys.~Rep.~{\bf 110}, 1 (1984).
\bibitem{kane} H.~E.~Haber and G.~L.~Kane, Phys.~Rep.~{\bf 117}, 75 (1985).
\bibitem{paige97} I.~Hinchliffe et al., Phys.~Rev.~D {\bf 55}, 5520 (1997);
                  F.~Paige, hep-ph/9801395, to appear in the proceedings of the
                  ''Workshop on Physics
                  at the First Muon Collider and the Front End of a Muon
                  Collider'', Batavia, IL, November 6--9, 1997.
\bibitem{gunion} J.~F.~Gunion, hep-ph/9802258,
to be published in the proceedings of ''Workshop on Physics at the First Muon 
Collider and
at the Front End of the Muon Collider'', Batavia, IL, 6--9 Nov 1997. 
\bibitem{Gunion96} J.~F.~Gunion et~al.,
  Proceedings of the 1996 DPF/DPB Summer Study on High-Energy Physics, 
  Snowmass, Colorado, June 25 -- July 12, 1996, eds. D.~G.~Cassel, 
                   L.~Trindle Gennari, R.~H.~Siemann, Vol.~2, p.~541.
\bibitem{haber} V.~Barger, M.~S.~Berger, J.~F.~Gunion, and T.~Han, 
                Phys.~Rep.~{\bf 286}, 1 (1997);
 H.~E.~Haber, talk given at the ''Workshop on Physics at the First Muon 
Collider and
at the Front End of the Muon Collider'', Batavia, IL, 6--9 Nov 1997, 
to be published in the proceedings.
\bibitem{Gunion86} J.~F.~Gunion, H.~E.~Haber, Nucl.~Phys.~{\bf 272}, 1 (1986).
\bibitem{higgshunter} J.~F.~Gunion, H.~E.~Haber, G.~L.~Kane, and S.~Dawson,
{\it The Higgs Hunter's Guide}, Addison-Wesley, 1990, and references therein.
\bibitem{Ellis83}
J.~Ellis and S.~Rudaz, Phys.~Lett.~B {\bf 128}, 248 (1983);
G.~Altarelli and R.~R\"uckl, Phys.~Lett.~B {\bf 144}, 126 (1984).
\bibitem{Bartl97} A.~Bartl, W.~Majerotto, and W.~Porod,
                  Z.~Phys.~C {\bf 64}, 499 (1994);
                  A.~Bartl et al., Z.~Phys.~C {\bf 73}, 469 (1997);
                  A.~Bartl et al., Z.~Phys.~C {\bf 76}, 549 (1997).
\bibitem{desy} E.~Accomando et al. (ECFA/DESY LC Physics Working Group), 
               DESY-97-100, (1997), submitted to Phys.~Rep.
\bibitem{ellis} J.~Ellis, G.~Ridolfi, and F.~Zwirner, Phys.~Lett.~B
               {\bf 257}, 83 (1991); Phys.~Lett.~B {\bf 262}, 477 (1991).
\bibitem{haber91} H.~E.~Haber and R.~Hempfling, Phys.~Rev.~Lett.~{\bf 66},
                  1815 (1991).
\bibitem{ross} L.~E.~Ib\'a\~{n}ez and G.~G.~Ross, Phys.~Lett.~B
               {\bf 110}, 215 (1982).
\bibitem{eberl}  H.~Eberl, A.~Bartl, W.~Majerotto,
                 Nucl.~Phys.~{\bf B472}, 481 (1996).
\bibitem{bartl96b} A.~Bartl, H.~Eberl, K.~Hidaka, T.~Kon, W.~Majerotto, and 
                   Y.~Yamada, Phys.~Lett.~B {\bf 389}, 538 (1996).
\bibitem{bartl97a} A.~Bartl, H.~Eberl, K.~Hidaka, T.~Kon, W.~Majerotto, and 
                   Y.~Yamada, Phys.~Lett.~B {\bf 402}, 303 (1997).
\bibitem{djouadi97} A.~Djouadi, J.~Kalinowski, P.~Ohmann, and 
                   P.~M.~Zerwas, Z.~Phys.~C {\bf 74}, 93 (1997).
\end{thebibliography}
\end{document}